\documentclass[conference]{IEEEtran}
\IEEEoverridecommandlockouts

\usepackage[utf8]{inputenc}
\usepackage{amsmath}
\usepackage{amsfonts}
\usepackage{amsthm}
\usepackage{graphicx}
\usepackage{hyperref}
\usepackage{booktabs}
\usepackage{multirow}
\usepackage{tabularx}
\usepackage{cite}
\usepackage{textcomp}
\usepackage{xcolor}
\usepackage{subcaption}
\usepackage{adjustbox}

\newtheorem{task}{Task}

\begin{document}

\title{Prompting for Performance: Exploring LLMs for Configuring Software%
\thanks{This work was supported through the Inria-Simula associate team RIPOST, the Inria D{\'{e}}fi LLM4Code project, and the European Commission through the projects AI4CCAM (Trustworthy AI for Connected, Cooperative Automated Mobility, grant agreement No 101076911), and AI4COPSEC (Boosting EU Copernicus Security and Maritime Monitoring with AI and Machine Learning (ML), grant agreement No 101190021).}%
}
\author{\IEEEauthorblockN{Helge Spieker\IEEEauthorrefmark{1}, Théo Matricon\IEEEauthorrefmark{2}, Nassim Belmecheri\IEEEauthorrefmark{1}, Jørn Eirik Betten\IEEEauthorrefmark{1}, 
Gauthier Le Bartz Lyan\IEEEauthorrefmark{2},\\  Heraldo Borges\IEEEauthorrefmark{2}, Quentin Mazouni\IEEEauthorrefmark{1}, Dennis Gross\IEEEauthorrefmark{1},
 Arnaud Gotlieb\IEEEauthorrefmark{1}, Mathieu Acher\IEEEauthorrefmark{2}}\\
\IEEEauthorblockA{\IEEEauthorrefmark{1}\textit{Simula Research Laboratory}, Oslo, Norway \\
\{helge,nassim,jorneirik,quentin,dennis,arnaud\}@simula.no}
\IEEEauthorblockA{\IEEEauthorrefmark{2}\textit{Univ Rennes, Inria, CNRS, IRISA}, 
Rennes, France \\
\{theo.matricon,gauthier.le-bartz-lyan,heraldo.pimenta-borges-filho,mathieu.acher\}@inria.fr}
}%

\maketitle

\begin{abstract}
Software systems usually provide numerous configuration options that can affect performance metrics such as execution time, memory usage, binary size, or bitrate. On the one hand, making informed decisions is challenging and requires domain expertise in options and their combinations. On the other hand, machine learning techniques can search vast configuration spaces, but with a high computational cost, since concrete executions of numerous configurations are required. In this exploratory study, we investigate whether large language models (LLMs) can assist in performance-oriented software configuration through prompts. We evaluate several LLMs on tasks including identifying relevant options, ranking configurations, and recommending performant configurations across various configurable systems, such as compilers, video encoders, and SAT solvers. Our preliminary results reveal both positive abilities and notable limitations: depending on the task and systems, LLMs can well align with expert knowledge, whereas hallucinations or superficial reasoning can emerge in other cases. These findings represent a first step toward systematic evaluations and the design of LLM-based solutions to assist with software configuration.
\end{abstract}


\section{Introduction}

Modern software systems (e.g., compilers, databases, ML pipelines) expose many configuration options impacting performance trade-offs like latency or memory usage.
Identifying performant configurations and key options is challenging due to combinatorial configuration spaces and complex, non-linear interactions, making analytical reasoning impractical~\cite{guo2013}.

ML-based approaches~\cite{DBLP:journals/jss/PereiraAMJBV21,guo2015,siegmund2015,nair2017,lu2022learning}, especially predictive models~\cite{Guo2017,lesoil_learning_2024}, assist in configuration but require empirical performance data. Collecting these data is often prohibitively expensive, limiting scalability and sometimes outweighing the performance gains.
Manual approaches relying on documentation, expert intuition, or community knowledge suffer from outdated/incomplete information, anecdotal evidence, and the need for domain expertise. Default configurations are often suboptimal~\cite{lesoil_input_2023,nair2018,Kaltenecker2020}. Consequently, manual configuration remains slow and error-prone.

Meanwhile, Large Language Models (LLMs) have emerged as versatile tools capable of performing a wide range of reasoning, generation, and comprehension tasks across diverse domains. Their ability to capture rich contextual information from natural language and code has led to widespread exploration in software engineering tasks -- from code completion to documentation generation~\cite{DBLP:journals/tosem/HouZLYWLLLGW24,10.1145/3712003}. This raises a natural question: \emph{Can LLMs assist with performance-aware software configuration?}
This leads us to a broader question: Could LLMs serve as a new \emph{modality} for supporting configuration tasks -- complementing or even replacing traditional sources such as documentation, expert knowledge, or empirical learning? Our hypothesis is that LLMs can provide meaningful guidance in navigating configuration spaces by leveraging their compressed knowledge and contextual reasoning.

In this work, we investigate how LLMs can help configure software systems for performance. 
Our study considers three core tasks: (1) identifying influential options, (2) ranking configurations by expected performance, and (3) recommending performant configurations. These tasks require domain knowledge, some reasoning, and contextual inference. %
We experiment with various LLMs and configurable systems. Our findings provide a nuanced picture: In some scenarios, LLMs align well with expert intuition and make sensible configuration choices; in others, they exhibit limitations such as hallucinations or oversimplified reasoning. These mixed results highlight both the potential and the current limitations of LLMs in this new application domain. %
Our work initiates the development of benchmarks for evaluating LLMs on software configuration tasks and reports on open challenges associated with their rigorous assessment.

\section{Performance Configuration in Software}
We use \texttt{x264}, a widely-used H.264 video encoder, to illustrate the challenges and opportunities of performance configuration. The system exposes dozens of options, including compression strategies, prediction modes, and optimization flags. Their effects on performance can be significant and non-obvious. %
Furthermore, recent studies show that other variability factors (e.g. workloads) can influence the software configuration space~\cite{lesoil_input_2023,muhlbauer2023impact}.
According to a survey by Pereira et al.~\cite{DBLP:journals/jss/PereiraAMJBV21}, \texttt{x264} is one of the most considered subject systems in the literature on the performance of configurable systems.

\begin{figure}[t]
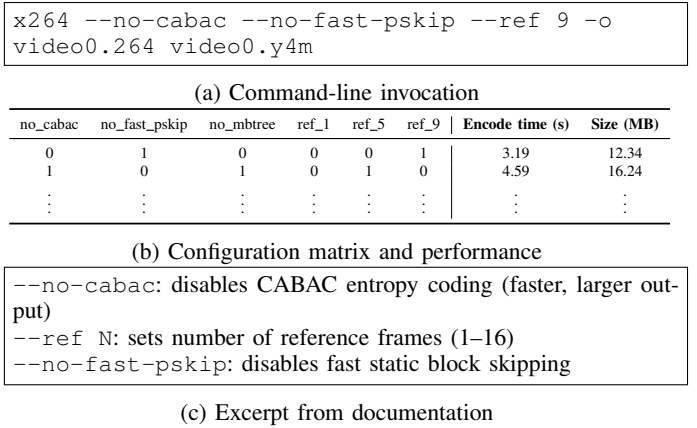

    \centering
    \begin{subfigure}[t]{\columnwidth}
        \centering
        \small
        \fbox{\begin{minipage}{\textwidth}
            \texttt{x264 --no-cabac --no-fast-pskip --ref 9 -o video0.264 video0.y4m}
        \end{minipage}}
        \caption{Command-line invocation}
    \end{subfigure}
    \\
    \begin{subfigure}[t]{\columnwidth}
        \centering
        \scriptsize
        \addtolength{\tabcolsep}{-1pt} 
        \adjustbox{max width=\columnwidth}{
        \begin{tabular}{cccccc|cc}
        \toprule
        no\_cabac & no\_fast\_pskip & no\_mbtree & ref\_1 & ref\_5 & ref\_9 & \textbf{Encode time (s)} & \textbf{Size (MB)}  \\
        \midrule
        0 & 1 & 0 & 0 & 0 & 1 & 3.19 & 12.34  \\
        1 & 0 & 1 & 0 & 1 & 0 & 4.59 & 16.24 \\
        \vdots & \vdots & \vdots & \vdots & \vdots & \vdots & \vdots & \vdots \\
        \bottomrule
        \end{tabular}}
        \addtolength{\tabcolsep}{1pt}
        \caption{Configuration matrix and performance}
    \end{subfigure}
    \\
    \begin{subfigure}[t]{\columnwidth}
        \centering
        \small
        \fbox{\begin{minipage}{\textwidth}
            \texttt{--no-cabac}: disables CABAC entropy coding (faster, larger output)\\
            \texttt{--ref N}: sets number of reference frames (1--16)\\
            \texttt{--no-fast-pskip}: disables fast static block skipping
        \end{minipage}}
        \caption{Excerpt from documentation}
    \end{subfigure}
    \caption{Different perspectives on \texttt{x264}}
    \label{fig:x264}
\end{figure}

\paragraph{Documentation vs. Learning} Developers and users often rely on documentation (Figure~\ref{fig:x264}c) to guide configuration decisions. Such descriptions provide an informal insight into the possible effects of individual options. However, they rarely account for interactions, hardware dependencies, or emergent behaviors~\cite{DBLP:conf/wosp/ValovPGFC17,jamshidi2017b,jamshidi2017a}. Another issue is that not all nonfunctional properties are documented. Moreover, the concrete, quantitative effect of the options is unclear.

A complementary strategy is to treat the system as a black box and measure its performance over sampled configurations (Figure~\ref{fig:x264}b). These measurements can be used to train predictive models~\cite{guo2013,Guo2017,DBLP:journals/jss/PereiraAMJBV21}, rank configurations~\cite{nair2017}, or optimize configurations under constraints~\cite{jamshidi2017a, bao2018}.  
The learning-based approach is often empirically effective: it can capture complex interactions among options, generalize across nearby configurations, and adapt to observed performance behaviors.
However, learning is costly~\cite{nair2017}. Each measurement incurs a real execution cost -- potentially minutes or hours of CPU time per configuration. Sampling strategies must balance exploration and budget constraints. Moreover, the quality of the model depends on the representativeness of the data, which requires careful experimental design.
In addition, applying learned models to new workloads or hardware often requires transfer learning or contextual adaptation, introducing additional cost~\cite{muhlbauer2020,martin:hal-03358817,jamshidi2017a}.

\paragraph{LLM for Performance Configuration}
This leads to a fundamental trade-off: relying on documentation is low-cost, but inaccurate, while learning-based models are powerful but costly. 
Within this spectrum, we hypothesize that LLMs can mitigate -- but not eliminate -- the measurement-vs-accuracy tension. These models have been trained on vast corpora (code, documentation, forums) and may encapsulate compressed domain knowledge queryable at low inference cost.
 Our key hypothesis is that LLMs can infer likely performance impacts of configuration options and reason about trade-offs without executing configurations. 
For instance, an LLM might recognize that enabling --no-cabac in x264 typically speeds up encoding at the expense of file size, even if this knowledge is not explicitly documented or measured in the immediate context. 
In a sense, LLMs could be viewed as a new \emph{modality} for configuration, in-between documentation, ML-based learning, and user knowledge. 
However, this potential needs validation. LLMs are not optimized for configuration reasoning, they may hallucinate and misunderstand performance semantics, or they can overgeneralize and rely on flawed/outdated training data~\cite{gallego2024, lian2024}, and, generally speaking, they lack guaranteed grounding in verifiable measurements or consistency over prompt variations.
The coexistence of promising capabilities and potential pitfalls in LLMs underscores the need for an exploratory study.

\subsubsection{Related Work}

Recent work has explored the use of LLMs to assist in software configuration tasks in a variety of domains, including Infrastructure as Code (IaC), cloud platforms, and system-level configuration~\cite{acher2023, Greiner2024, lian2024, zheng2024, chan2024, Rozire2023, Wang2024, xie2025, Goyal2025, Zhang2024}. The applications span the Linux kernel settings~\cite{Yang2025}, validating the configuration files~\cite{lian2024a, gallego2024}, and reasoning about constraints~\cite{michailidis2024, lira2024}. LLMs have also been used to model variability and generate feature models~\cite{Galindo2023}. Greiner et al.~\cite{Greiner2024} propose levels of generative AI support for software product line engineering (see also Section~\ref{sec:conclusion}).
Integrating LLMs into configuration workflows offers benefits such as increased automation, reduced manual errors, and improved user experience through natural language~\cite{lian2024,chan2024}. 
To our knowledge, our study is the first to consider the use of LLMs for \emph{performance-oriented} configuration in highly configurable systems.

\section{Evaluating Language Models for Software Configuration}

We will first discuss our proposed task selection to assess the suitability of LLMs for software configuration.
We are proposing a suite of three evaluation tasks that address different ways of supporting the software configuration problem: 1) configuration knowledge, 2) configuration selection, and 3) configuration recommendation.
Each of these tasks addresses a specific configuration challenge with a varying degree of user assistance and offers a different perspective on LLM capabilities in software configuration.

\begin{task}[Configuration Knowledge]
Given a system description and a target metric to optimize, the LLM should give a ranking of the most influential configuration options to affect the target metric.
The task is evaluated by comparing the ranking against a groundtruth ranking, e.g., through Kendall's rank correlation coefficient \(\tau\), or, if a groundtruth is unavailable, the consistency among multiple LLMs can be evaluated, even though that is less informative regarding the overall correctness of the ranking.
\end{task}

\textit{Configuration Challenge} 
The LLM is used as an assistant for the user to help with the construction of a custom configuration. 
It should point the user towards the most influential configuration options for their chosen target metric.
The user can then consider the options and learn more about them, or test their effect for his specific setting.

\textit{LLM Challenge} 
Configuration knowledge is a generic task without dependence on an exact input. Given a particularly popular system, the LLM might have encountered large amounts of documentation, forum discussions, or articles on the proper configuration of the system, and should be able to produce a reasonable ranking of the most influential configuration options.

\begin{task}[Configuration Selection]
    Given a system description, an input description, a target metric to optimize, and a set of $n$ configurations, the LLM should rank the configurations according to their performance for the target metric.
    The task is evaluated by the rank of the best performing configuration, respectively, the average rank over multiple iterations.
\end{task}

\textit{Configuration Challenge}
A user might have a collection of configurations that they are equally satisfied with in terms of other constraints or metrics, but need assistance in selecting the best one from this pool for a specific input.

\textit{LLM Challenge}
This task is particularly challenging as it requires analyzing given configurations with all options and their interactions, and assessing their effect for a given input.
It requires dedicated knowledge about the configuration space.

\begin{task}[Configuration Recommendation]
    Given a system description, an input description, and a target metric to optimize, the LLM should provide an executable configuration.
    The task is evaluated by the achieved performance. If the optimal performance is unavailable -- which is commonly the case in software configuration --, it requires the availability of strong baselines or competitor LLMs for a relative evaluation.
\end{task}

\textit{Configuration Challenge}
The user fully delegates the task of finding a good configuration to the LLM.

\textit{LLM Challenge}
This task provides the LLM the most freedom and the capability to make use of any knowledge that is available, either from its training data, as part of the prompt, or through external sources available to the LLM.

We propose these three tasks as they cover different degrees of user assistance, ranging from supporting a user to learn about the system and construct their own configuration, to assisting in deciding between externally provided configurations, e.g., from historical data or domain expertise, to fully taking over the task by recommending a specific configuration.
All of these tasks face a common challenge, which is the availability of groundtruth data for their evaluation. While configuration selection is inherently referring only to the relative performance of the configurations, groundtruth data can be collected from performing actual measurements, and existing datasets can be used. For the other tasks, determining the actual optimum results is more difficult, especially for complex real-world systems. Here, we have to rely on approximations of the groundtruth (Configuration Knowledge) or relatively comparison to baselines or other LLMs (Configuration Recommendation). 

In the following, we will present an implementation and evaluation of these tasks on a selection of highly configurable systems with a few current LLMs.

\section{Evaluation Results}

We perform an evaluation of LLMs with the three previously introduced tasks, that test different ways to use LLMs for software configuration: Configuration Knowledge, Configuration Selection, and Configuration Recommendation.

\subsection{Experimental Setup}

\subsubsection{Large Language Models}
Our goal in this study is not to benchmark which LLMs are most capable of aiding a user in software configuration, but to evaluate whether general, off-the-shelf LLMs are capable of being used for software configuration at all -- the progress in the LLM space is rapid and any benchmark can only form a snapshot, but if we find LLMs to be capable to solve the tasks already, then future LLMs will very likely have the same ability.
Therefore, we focus foremost on large commercial LLMs, which are currently at the forefront of LLM capabilities, but without being exhaustive, and while being aware that there are frequent releases of models with claimed superior performance. This evaluation is therefore a snapshot in time to identify the current capabilities.
Specifically, we use Anthropic's Claude 3.7 Sonnet, OpenAI GPT-4o (2024-11-20), DeepSeek Reasoner/R1, and Meta's Llama 4 Maverick (400B parameters, 8-bit floating point weights) as an openly available model.
For tasks 2 and 3, we additionally consider OpenAI GPT-4.1. 

\subsubsection{Configurable Systems}
In addition to the x264 video encoder, additional systems are considered.
The selection of systems is based on the dataset created by Lesoil et al.~\cite{lesoil_input_2023}, who measured the performance of eight systems for many inputs and configurations.
For all tasks, we rely on the data from this dataset, including the input properties to describe the input and the collected measurements as ground-truth data for the ranking of configurations.
Our code, prompt examples, and results are  available online\footnote{\url{https://anonymous.4open.science/r/promptforperf-ictai}}.

\subsubsection{Prompt Design}
A key aspect in the study of LLMs is the design of the prompts used to interact with them.
This prompt engineering is both model-\cite{polo2024efficient} and task-specific~\cite{vatsal2024survey} and, while it can make a difference in LLM performance, it is not reasonable to assume an end-user would pour extensive efforts into prompt optimization for their specific inputs.
In the context of our evaluation, we chose a middle ground between the most naive prompt and a highly optimized prompt.
We have made the decision not to choose any particular prompt optimization strategy, e.g., DSPy~\cite{khattab2024dspy}, involvement of external sources, like, for example, retrieval-augmented generation (RAG)~\cite{DBLP:conf/nips/LewisPPPKGKLYR020}, or particular prompting strategies like Chain-of-Thought~\cite{wei2022chain}, which is already implicitly done in reasoning models like the used DeepSeek R1.
Instead, we are using a one-shot prompting setting for all our prompts without any iteration, except for its internal reasoning, if available.
We have revised prompts empirically through preliminary experiments.

\subsection{Task 1: Configuration Knowledge}
In the first task, we study the ability of LLMs to rank different configurations according to their impact on a specific metric.
In order to do that, we prompt LLMs to rank configuration options to improve the bitrate of the encoded video. Then we measure the agreement between repetitions of the same prompt for the same LLM. We then measure the agreement between the LLMs. Finally, we evaluate how the LLM's ranking relates to real empirical data.

\subsubsection{Setup}
We selected the x264 system and asked each of the four LLMs the same prompt 20 times.
The prompt for task 1 is shown in \autoref{fig:prompt_task1}.
We parsed the answers of the LLMs with a Python script and, using some expert knowledge, merged different naming conventions for the LLMs' answers.

\begin{figure}
    \centering
    \begin{subfigure}[t]{\columnwidth}
        \centering
        \small
        \fbox{\begin{minipage}{\textwidth}
        You are a leading expert in video encoding with deep knowledge of x264 encoder performance optimizations.\\
            Your task is to return a x264 command to maximize the bitrate of the encoded video.
    Please provide a ranked list (from most to least important) of configuration options for the x264 encoder that impact bitrate positively. For each option, include a brief one-sentence explanation of its effect on bitrate. Focus exclusively on bitrate and do not consider things like CPU usage, threading, and algorithmic optimizations.
        \end{minipage}}
    \end{subfigure}\\
    \caption{Prompt schema for task 1.}
    \label{fig:prompt_task1}
\end{figure}

\subsubsection{Option Ranking}
\begin{figure}
    \centering
    \includegraphics[width=0.95\linewidth]{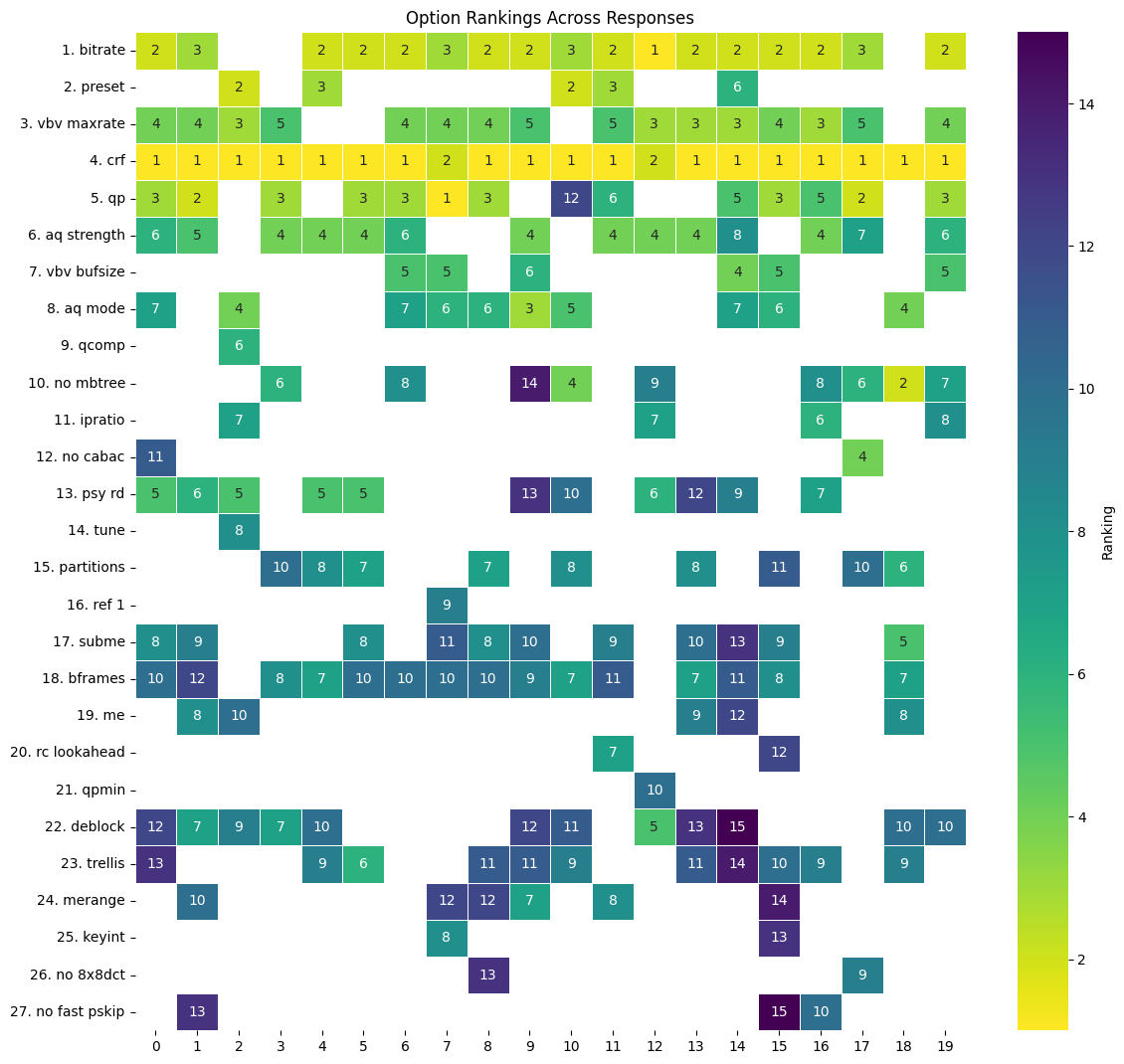}
    \caption{Ranking of Options for DeepSeek R1 based on their impact on bitrate}
    \label{fig:ranking_deepseek}
\end{figure}

\begin{figure}
    \centering
    \includegraphics[width=0.95\linewidth]{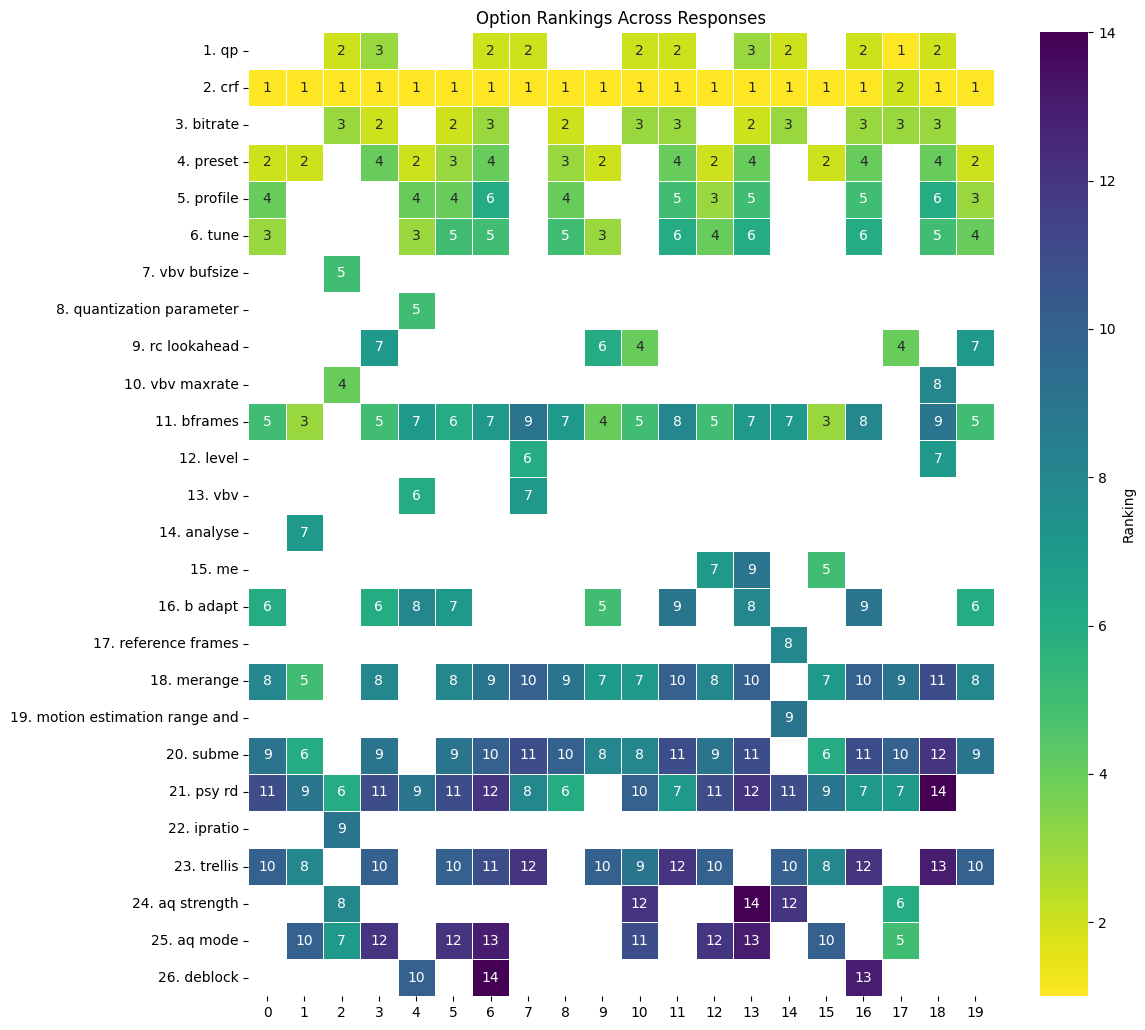}
    \caption{Ranking of Options for Llama 4 based on their impact on bitrate}
    \label{fig:ranking_llama}
\end{figure}

We only show the ranking for two LLMs to save space; on \autoref{fig:ranking_deepseek} for DeepSeek and on \autoref{fig:ranking_llama} for Llama 4.
The behaviors of the other two models are more or less identical to the two shown here, a notable difference is the number of different `top' options offered by Claude which amount to 46.
All LLMs seem to provide a consistent top five list of options, whereas the rest of the recommendations seem more random.


\subsubsection{Stability of Rankings for the same LLM in all answers}
To measure the stability of the LLMs' answers, we repeated the same propositions multiple times and measured the agreement between these different rankings using the Spearman correlation coefficient. 

The results are reported in \autoref{tab:spearman_results}.
GPT-4o is more stable in its answers than other LLMS. DeepSeek has a consistent top 5 options a bit behind, then Llama 4. DeepSeek is less consistent in its top 5 and top 3 than in its top 10.
Overall, DeepSeek is more chaotic generally, but other LLMs seem to follow the same trend as the number of answers increases.

\begin{table}[t]
\small
\centering
\caption{Task 1 - Stability of Rankings: Average Spearman Correlation for Different Models}
\label{tab:spearman_results}
\begin{tabular}{lccc}
\toprule
\textbf{Model} & \textbf{Top 3} & \textbf{Top 5} & \textbf{Top 10} \\
\midrule
Claude 3.7 Sonnet & 0.473 & 0.496 & 0.550 \\
DeepSeek R1 & 0.695 & 0.625 & 0.499 \\
Llama 4 Maverick & 0.624 & 0.573 & 0.534 \\
GPT-4o & \textbf{0.738} & \textbf{0.710} & \textbf{0.699}  \\
\bottomrule
\end{tabular}

\end{table}

\subsubsection{Agreement of Rankings across LLMs and real data}
We measure the agreement of the ranking between LLMs by comparing their mean answer ranks using the mentioned metric.
We sampled different configurations taken from the data generated by the other tasks in order to have real data about the options' impact. 
Due to the lack of a groundtruth ranking, we rely on statistical tests instead to evaluate an option's impact and ranking~\cite{Grochtmann2004StatisticalOracle,Barr2014OracleSurvey}.
To compare two options, A and B, we compare the bitrate of the configuration with option A and without option B, with the one of option B and without option A.
This gives us paired data ideal for a Wilcoxon signed rank test. However, the data was missing for many configurations, so we ran x264 on the same inputs but with different configurations. This led us to only consider the top 3 options ignoring bitrate, the most stable, which are: preset, qp and crf. The latter two with values 0 and preset with the veryslow flag.
Note that while those three were the top options for all of them, they were not always present in the ranking of top options by the LLMs.

Most LLMs rank crf as the most impactful option so we ran the statistical tests with this hypothesis, then qp is ranked after preset, but it depends on the LLM asked, so for this case we split the data in two. 
We ran one-sided tests: crf is better than preset with effect size 1 and p-value $< 10^{-14}$; crf is better than qp with effect size 0.56 and p-value $< 10^{-7}$;  preset is not better than qp with effect size -1 and p-value 1 on the first half of the data so we use the other half of the data to test the other hypothesis and qp is better than preset with effect size 1 and p-value $< 10^{-7}$.

It seems most LLMs get the correct top options, but the high variance in their rankings make them unreliable, it is likely that multiple samples should be taken to produce an accurate ranking, even then while the top options seem to agree with reality it is not clear in which order they are most important; indeed most LLMs put preset as more important than qp while it is the opposite.

\subsubsection{Replication with encoding time}
 We did the same set of experiments with encoding time and found similar behaviors (see online supplementary). First, LLMs consistently produce a top 5 ranking, showing internal coherence. Second, the outputs beyond top 5 become more chaotic. 
For encoding time, GPT-4o is the most chaotic, while DeepSeek is the most stable -- unlike what we observed for bitrate -- highlighting that models' behavior vary \emph{across metrics}.

\subsubsection{Conclusion}
Overall, most LLMs given enough samples find the most relevant options. However, the exact ranking is unreliable, suggesting a strategy of sampling multiple rankings and aggregating the results to get the most relevant options.
The top 3 or top 5 relevant options are reliably identified, which are the most impactful options for practitioners. However, beyond the few most relevant options, the rankings provided by the LLMs are unreliable.

\subsection{Task 2: Configuration Selection}

\begin{table*}[t]
\small
\centering
\caption{Task 2 - Selection: Rank of the best configuration (1: best, \(n\): worst)  for \(n=2\) and \(n=5\). DS R1: DeepSeek R1.}
\label{tab:task2}
\begin{tabular}{l rrrrr | rrrrr}
\toprule
 & \multicolumn{5}{c}{\(n=2\)} & \multicolumn{5}{c}{\(n=5\)} \\
\cmidrule(lr){2-6} \cmidrule(lr){7-11} %
System & Claude 3.7 & DS R1 & Llama 4 & GPT-4.1 & GPT-4o 
      & Claude 3.7 & DS R1 & Llama 4 & GPT-4.1 & GPT-4o \\
\midrule
gcc & \textbf{1.38} & 1.38 & 1.39 & 1.56 & 1.51 & 3.42 & \textbf{2.30} & 2.58 & 2.55 & 2.81 \\
imagemagick & \textbf{1.50} & 1.58 & 1.70 & 1.52 & 1.62 & \textbf{3.35} & 3.89 & 4.07 & 3.49 & 3.55 \\
lingeling & \textbf{1.44} & 1.53 & 1.45 & 1.47 & 1.54 & \textbf{2.99} & 3.04 & 3.10 & 3.00 & 3.28 \\
nodejs & 1.39 & 1.39 & 1.40 & 1.41 & \textbf{1.36} & 2.47 & 2.48 & 2.53 & 2.56 & \textbf{2.43} \\
poppler & 1.42 & 1.43 & 1.51 & \textbf{1.42} & 1.50 & 2.81 & 2.79 & 2.94 & \textbf{2.54} & 2.79 \\
sqlite & 1.58 & 1.48 & 1.51 & \textbf{1.46} & 1.57 & 2.81 & 3.25 & 3.10 & 3.15 & \textbf{2.70} \\
x264 & 1.61 & \textbf{1.36} & 1.57 & 1.60 & 1.67 & 4.24 & \textbf{1.60} & 4.27 & 4.04 & 4.30 \\
xz & 1.36 & 1.51 & 1.46 & \textbf{1.32} & 1.35 & 2.61 & 2.64 & 2.95 & 2.73 & \textbf{2.57} \\
\midrule
Average & 1.46 & 1.46 & 1.50 & 1.47 & 1.52 & 3.09 & \textbf{2.75} & 3.19 & 3.01 & 3.05 \\
\bottomrule
\end{tabular}
\end{table*}

The second case study tests the capabilities of LLMs to select well-performing configurations from a set of given alternatives.
In the experiment, the LLM is prompted with a description of the input and $n$ configurations.
It is asked to rank those configurations according to their expected performance for a given performance metric, e.g., to reduce the size of the compiled program in GCC.

\subsubsection{Setup}
We consider two variants of this experiment.
In the first, we choose $n=2$, i.e., we ask the LLM to select the better of two configurations. 
In the second, we choose $n=5$, asking for a ranking of the configurations.
In both cases, we evaluate the average rank of the correct configuration, i.e., the optimal result would be 1, the worst is $n$.
In each experiment, we randomly select 100 inputs and $n$ configurations per input.

The prompt for task 2 is shown in Figure~\ref{fig:prompt-shared} and Figure~\ref{fig:prompt-task2}.
Both task 2 and task 3 have the same general prompt structure, with a shared preamble to prime the LLM on its task and to provide input-specific information.
Examples of prompts for all systems are available in the online material.

\begin{figure}
    \centering
    \begin{subfigure}[t]{\columnwidth}
        \centering
        \small
        \fbox{\begin{minipage}{\textwidth}
            You are a leading expert in video encoding with deep knowledge of x264 encoder performance optimizations.\\
            Your task is to return a x264 command to maximize the bitrate of the encoded video.\\
            (Optional context included here (Task 3 only))\\
            I have a raw mkv file with the following properties:\\
            - property name: property value\\
            - \dots
        \end{minipage}}
        \caption{Shared initial prompt for Task 2 and 3 on the example of x264\label{fig:prompt-shared}}
    \end{subfigure}\\
    \begin{subfigure}[t]{\columnwidth}
        \centering
        \small
        \fbox{\begin{minipage}{\textwidth}
    Please rank the following configurations from best to worst performance to maximize the bitrate.
    I'm providing all configuration options I set, any unspecified option is left at its default value.
    Return the ranking of the IDs of the configurations in a \textbackslash boxed\{\} environment separated by commas, e.g. \texttt{\textbackslash boxed\{1,2,3,4,5,6\}}. No explanations.
        \end{minipage}}
        \caption{Prompt continuation for Task 2\label{fig:prompt-task2}}
    \end{subfigure}\\
    \begin{subfigure}[t]{\columnwidth}
        \centering
        \small
        \fbox{\begin{minipage}{\textwidth}
    The input is a raw mkv file. 
    What x264 configuration options maximize the bitrate? 
    Return the x264 command in a \textbackslash boxed\{\} environment. Use \(<\)OUTPUT\(>\) as the output file and \(<\)INPUT\(>\) as the input file, e.g., \texttt{\textbackslash boxed\{x264 <INPUT> -o <OUTPUT>\}}. You are not allowed to use the --bitrate parameter.
        \end{minipage}}
        \caption{Prompt continuation for Task 3\label{fig:prompt-task3}}
    \end{subfigure}
    \caption{Prompt schema for tasks 2 and 3 with a shared prompt preamble. All other systems are described similarly.}
    \label{fig:prompts}
\end{figure}

\subsubsection{Results}
The left part of Table~\ref{tab:task2} shows the results for the $n=2$ case.
None of the selected LLMs show consistently strong performance in selecting the better of the two configurations.
Although individual LLMs show good performance for some systems, none of them performs on average much better than random choice.

The right part of Table~\ref{tab:task2} shows the result for $n=5$, which is similar to the previous case.  
Since there are multiple configuration choices, the LLM was prompted to rank them in order of their performance. The results show that the average rank of the best configuration is not close to the top position. Again, Claude 3.7 and DeepSeek R1 show the best performance, but still have shortcomings in solving the task.

What could be the reason for this poor performance?
First of all, it is important to note that we prompt the LLM without additional information about the system, that is, we test the LLM's internal knowledge of configuration options, their interaction, and their effect on the given input, which is a difficult task.
Second, the ability to select the correct configuration is restricted by the prompt, and especially the input properties given.
Here, we rely on the set of input properties given in the dataset~\cite{lesoil_input_2023}, while we acknowledge that they might not be the most relevant for our task.
Third, this task has a high difficulty.
Solving it correctly requires a structured internal representation of how configuration parameters interact with each other to influence performance metrics under varying inputs (and their potentially limited description).
In the context of this study, the configuration selection is probably the most ambitious task and the one where a very strong LLM performance would have been the most surprising.
Still, it shows the limitations of current LLMs and the development opportunities for future models.

\subsection{Task 3: Configuration Recommendation}

The third task addresses whether LLMs can recommend an entire configuration for a given input and performance optimization goal, on the example of the x264 video encoder.

\subsubsection{Setup}
We sample 30 videos from the YouTube UGC dataset~\cite{DBLP:conf/mmsp/WangIA19} covering different video categories and resolutions and query the LLM to return an x264 command that maximizes the bitrate of the encoded video.
Each input is described by the features provided in~\cite{lesoil_input_2023}.
We consider two variants, one in which the prompt is enhanced with additional context, that is, the x264 documentation as provided by \verb|x264 --fullhelp|, and one without any additional context.
The encoding is performed on an M2 MacBook Pro with 32\, GB RAM and x264 version 0.164.3191M.

Each LLM is queried five times per video to account for nondeterminism in the response and to test for stability of the results.
Performance is evaluated by comparing the relative performance of the recommended configuration on the encoding of the video with the default configuration, i.e., without any additional parameters set.
We consider average performance over all iterations, as well as only the best and worst results over the five iterations per input.
As additional baselines, we have considered the x264 presets that configure a developer-chosen set of parameters to trade-off encoding speed and quality (\verb|--preset| = \{\verb|placebo|, \verb|veryslow|, \verb|slow|, \verb|fast|, \verb|ultrafast|\}).
However, they are not competitive to the LLM-provided configurations (the average relative performance is 2.3), and we do not report their results separately.
Additionally, we have compared the performance measures from the dataset in \cite{lesoil_input_2023}, which uses the same videos but has only a restrictive set of configuration options (as discussed above), as an additional baseline.
Again, the results were significantly outperformed by all LLM models, too, such that we do not further discuss them. 
Still, this shows that there is no pollution of the LLM training data from the dataset of~\cite{lesoil_input_2023}, otherwise the models would match the dataset performance. %

The task-specific prompt for task 3 is shown in Figure~\ref{fig:prompt-task3} with the shared preamble in Figure~\ref{fig:prompt-shared}.

\subsubsection{Results}
\begin{figure}
    \centering
    \includegraphics[width=\columnwidth]{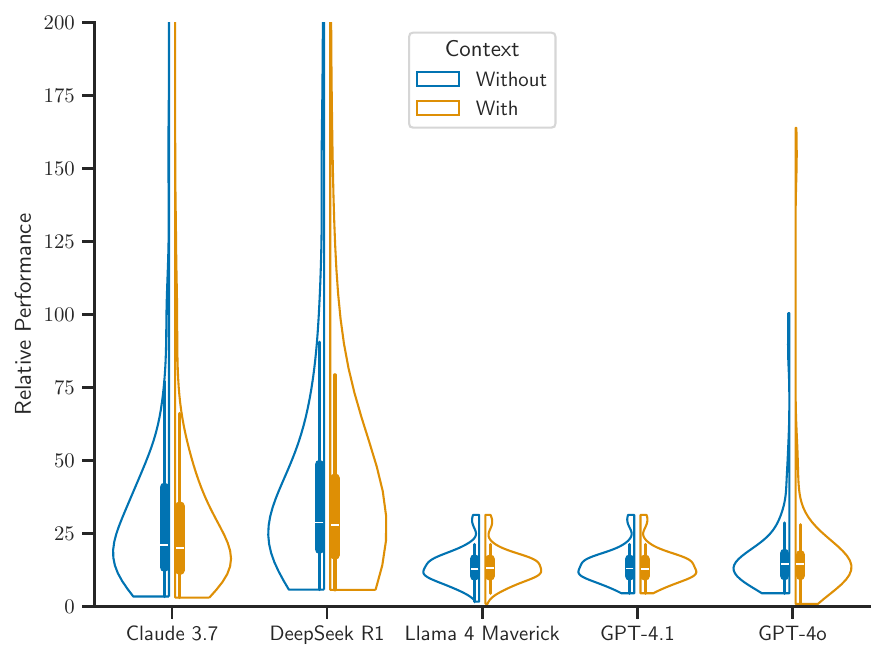}
    \caption{Task 3 - Recommendation of x264 configurations: Relative bitrate with recommended over default configuration. For better readability, we cut the graph at 200.}
    \label{fig:task3}
\end{figure}

The results are shown in Figure~\ref{fig:task3}. %
All LLMs are capable of selecting feasible and strong configurations that outperform the default configuration with statistical significance, as tested by a one-sided Wilcoxon signed-rank test.
However, the performance among the LLMs varies with DeepSeek R1 and Claude 3.7 showing the best performance, although with some variety between best and worst results.
Note that the graph is cut at a relative performance of 200; there are single configurations for Claude 3.7 and DeepSeek R1 that reach 270, respectively 840, but we decided on a better visibility of the main results.

Adding x264 help as context could be expected to help the configuration recommendation, as it provides additional information about available options and their meaning.
Here, we observe negative results.
Adding context does not change the results for Llama 4 or the GPT models.
Although the results for Claude 3.7 and DeepSeek R1 vary between the two settings with some higher outliers, but an overall slightly lower mean, the differences are not statistically significant (p-value $\approx 0.5$) and are likely due to model nondeterminism.
 Some hypotheses can account for the absence of measurable gains when the x264 help text is appended. A first hypothesis is that the documentation's content is already encoded in the models’ pre-training corpora, so re-presenting it contributes no additional information or better attention. A second hypothesis is that effective configuration should consider empirical interactions among both encoder options, hardware characteristics, and input properties. Such relationships are absent from any static manual and thus unrecoverable without runtime observations. 

\subsubsection{Overall conclusion}
Any LLM considered in the study is capable of providing valid and performant configurations, outperforming the default; however, with quality differences.

\section{Threats to Validity}

There are some threats to the validity of the presented results:
One comes from the selection of systems and datasets.
These systems are a sample of diverse applications, although not representative of the entire space of configurable software systems.
We select these specific systems from~\cite{lesoil_input_2023} to reuse the collection of input features, configurations, and measurements as the basis of our experiments.
The available input features might affect the effectiveness of the LLM in making configuration decisions.
A similar threat also applies to the LLMs, which are a small subset.
However, we did not aim to perform an exhaustive benchmark and selected the currently most powerful LLMs.
Finally, another threat lies in the construction of the tasks and especially in how we prompt the LLM and how we interact with it, e.g., not allowing tool use, external sources, or similar.
Prompt design can be crucial for LLM usability, and we varied the prompt in preliminary studies, but have not yet performed a structured evaluation. 
We acknowledge these threats -- after all, LLMs are highly configurable systems themselves -- and agree that a broader and more standardized evaluation will be fruitful future work.

\section{Conclusion}
\label{sec:conclusion}
This exploratory study investigated the potential of large language models (LLMs) to support performance-aware software configuration. 
We evaluated several state-of-the-art models on three novel tasks: identifying influential configuration options (Task 1), ranking candidate configurations (Task 2), and recommending performant configurations from scratch (Task 3). 
Our results paint a nuanced picture: while LLMs show clear promise for this domain, they also exhibit significant limitations, positioning them as powerful but not yet fully reliable assistants.
To the best of our knowledge, this is the first comprehensive evaluation of LLMs for performance-oriented configuration in highly configurable software systems.

Our findings vary distinctly across the tasks.
In \textbf{Task 1} (Knowledge), LLMs demonstrated a strong ability to identify globally important configuration options, being relatively stable among the top 3 options. Compared to real data, they do not always return the correct order, even for the top options, which are sometimes not suggested by the LLMs, indicating that precautions should be taken when considering only a few options. This suggests LLMs can act as effective, low-cost knowledge bases for identifying high-impact parameters, given they are sampled multiple times and the order of the top options is ignored.

In \textbf{Task 2} (Selection), LLMs were asked to select the best-performing configuration from a small set. 
Here, the LLMs failed to perform better than random choice. This highlights a critical limitation: LLMs currently lack the fine-grained, quantitative reasoning required to discern subtle performance differences between similar, valid configurations.

Finally, in \textbf{Task 3} (Recommendation), LLMs showed significant promise by generating novel configurations that consistently outperformed system defaults by a large margin, however, with variations among the tested LLMs. 
This indicates their potential as powerful assistants for generating strong ``first-guess'' configurations, even if their output quality varies.

Taken together, these results confirm that LLMs are \emph{not yet reliable standalone tools} for all performance configuration tasks.
Instead, they represent a promising new modality that complements existing approaches. 
They show great potential at high-level knowledge retrieval (Task 1) and creative generation (Task 3), but show their limitations on precise, comparative reasoning (Task 2). 
This makes them valuable, low-cost assistants for early-stage exploration or users seeking a better-than-default starting point, rather than fully automated optimizers.

These findings emphasize off-the-shelf results without additional, involved setups of the LLMs, like finetuning, extensive prompt optimization, or retrieval-augmented generation (RAG). These are all directions for future work, going into the design of configuration approaches involving LLMs rather than testing pure LLM capabilities.

Our study also contributes to building the foundation for more systematic evaluation: we propose and demonstrate task definitions, provide concrete metrics to assess LLM behavior, such as ranking stability, alignment with empirical measurements, and prompt repeatability, and offer data sets and experimental setups that can be reused and extended in future work.
We hope our open benchmark and evaluation protocol will help others reproduce and extend our work.
These contributions also highlight the need for standardized benchmarks that evaluate LLMs in a variety of configuration tasks, systems, and performance contexts. 
Still, the design of rigorous benchmarks is not trivial: benchmarking should account for variability across inputs, hardware, and environments; it should test robustness, hallucination resistance, and sensitivity to prompt structure and context. 
Designing and developing these standardized, reproducible, and extensible evaluation benchmarks will be key to understanding and improving LLMs in this domain.

In the broader context of generative AI for software variability, our work is positioned between Level~2 (basic support) and Level~3 (advanced support) in the Greiner et al. framework~\cite{Greiner2024}, which considers dimensions such as automation, output quality, and task complexity. Given the capabilities observed -- particularly in Task~1 (Knowledge) and 3 (Recommendation) -- we assert that LLMs are already offering valuable advanced support for certain configuration scenarios, although with important limitations in precision and reliability that preclude full automation.

In summary, LLMs offer a promising new opportunity for navigating configuration complexity when it comes to performance. 
Successfully integrating them into developer workflows, by leveraging their strengths while mitigating their weaknesses, presents a key challenge and an opportunity for the software engineering and AI communities.

\bibliographystyle{IEEEtran}
\bibliography{refs_ictai}
\end{document}